\documentclass[12pt]{iopart}

\usepackage{graphicx}
\usepackage{wrapfig}
\usepackage{appendix}
\usepackage{subfigure}

\begin{document}

\title{Ancilla-based quantum simulation}
\author{Katherine L. Brown$^1$, Suvabrata De$^1$,Vivien M. Kendon$^1$,William J. Munro$^{2,3}$}
\address{$^1$Department of Physics and Astronomy, University of Leeds, Leeds, LS2 9JT, UK}
\address{$^2$NTT Basic Research Laboratories, NTT Corporation, 3-1 Morinosato-Wakamiya, Atsugi-shi, Kanagawa-ken 243-0198, Japan}   
\address{$^3$National Institute of Informatics, 2-1-2 Hitotsubashi, Chiyoda-ku, Tokyo 101-8430, Japan}
\date{\today}

\begin{abstract}    
We consider simulating the BCS Hamiltonian, a model of low temperature superconductivity, on a quantum computer. In particular we consider conducting the simulation on the qubus quantum computer, which uses a continuous variable ancilla to generate interactions between qubits. We demonstrate an $O(N^3)$ improvement over previous work conducted on an NMR computer [PRL \textbf{89} 057904 (2002) \& PRL \textbf{97} 050504 (2006)] for the nearest neighbour and completely general cases. We then go on to show methods to minimise the number of operations needed per time step using the qubus in three cases; a completely general case, a case of exponentially decaying interactions and the case of fixed range interactions. We make these results controlled on an ancilla qubit so that we can apply the phase estimation algorithm, and hence show that when $N\ge5$, our qubus simulation requires significantly less operations that a similar simulation conducted on an NMR computer. 
\end{abstract}

\section{Introduction}
Quantum computing is a field of research which exploits the quantum behaviour of systems to get advantages over standard classical digital computing. The most famous example is Shor's algorithm, which gives an exponential improvement in factoring numbers compared to the best known classical algorithm \cite{Shor1997}. However, even if we ignore the need for fault tolerance, such algorithms require thousands of qubits to obtain results inaccessible to classical computers and are therefore unlikely to be useful in the near future. A more accessible problem which is comparably hard to perform on a classical computer is simulating quantum systems. The idea of using quantum computers to simulate quantum systems was first proposed in 1982 by Feynman \cite{Feynman1982}, and further developed by Lloyd \cite{Lloyd1996} in 1996. Quantum simulation on a quantum computer would provide us with efficient algorithms to obtain: ground states of molecules \cite{Aspuru-Guzik2005}; details about bond formation \cite{Smirnov2007}; and eigenvalues and eigenvectors \cite{Abrams1999} of many-body quantum Hamiltonians; for a recent review, see Brown et al.~\cite{Brown2010}. One of the largest systems of qubits which has been simulated in full generality on a classical computer consists of only 36 qubits \cite{deraedt06a}. By taking advantage of the symmetries that occur in many quantum systems it is possible to simulate larger systems on a classical computer \cite{verstraete04a}. While 36 qubits is still a significant increase on what is currently available, it is an order of magnitude smaller than the break even point for many other quantum algorithms. Even with only tens of qubits it is possible to obtain interesting results that would be computationally intensive to obtain classically, especially when considering the general case scenarios. With more advanced and highly controlled quantum computers it will be possible to get savings on important simulations such as chemical reactions; so not only is quantum simulation a test bed for quantum computing, it will also significantly increase our understanding of many scientific problems. 

One particular system of interest for an early simulation is the BCS Hamiltonian, a model of superconductivity formulated by Bardeen, Copper and Schrieffer \cite{BCS57}. Since the BCS model of superconductivity is still poorly understood, a quantum simulation of the system could help rectify this situation. The same Hamiltonian is also used to describe pairing in nuclear physics, where it is called the pairing force Hamiltonian \cite{Richardson1963}. Pairing Hamiltonians are used to describe many processes in condensed matter physics and therefore a technique for simulating the BCS Hamiltonian should be adaptable to many other purposes.  The BCS Hamiltonian, $H_{BCS}$ is
\begin{equation}
H_{\mathrm{BCS}} = \sum^{N}_{m=1} \frac{\epsilon_{m}}{2} (n_{m}^{F} + n_{-m}^{F}) + \sum^{N}_{m,l=1} V_{ml}C_{m}^{\dagger} C_{-m}^{\dagger} C_{-l} C_{l} 
\label{BCSpair}
\end{equation}
where $C_{m}^{\dagger}$ and $C_{m}$ are the Fermionic creation and annihilation operators, $n_{\pm m}^{F} = C_{\pm m}^{\dagger} C_{\pm m}$ are the number operators, $N$ is an effective state number that represents the number of modes to be simulated, $V_{ml}$ is the interaction potential, and $\epsilon_{m}$ is the on-site energy of a pair in mode $m$. Pairs of fermions have quantum numbers $m$ and $-m$ where pairs have equal energies but opposite momentum and spin, $m=(\mathbf{p},\uparrow)$ and $-m=(\mathbf{-p},\downarrow)$.

The energy gap between the ground and first excited state of the BCS Hamiltonian is a non-perturbative function, this means that the Hamiltonian needs to be exactly diagonalised to obtain an accurate value for the energy gap. The non-pertubative nature of the Hamiltonian also means that to get an accurate value for the energy gap it is important to consider information from all interactions near the Fermi surface including long range interactions \cite{Anderson1958}. To get a solution for the energy gap, Bardeen el al.~\cite{BCS57} took $V_{ml} = -V$ for states near the Fermi surface, and 0 elsewhere. This approximation works for most metallic superconductors, but there are some cases where it is inaccurate. In these cases it would be useful to perform a simulation with generalised $V_{ml}$. While there have been attempts to do this classically, many classical approximation methods either don't take into account these long range interactions, or require them to obey very specific rules \cite{Mastellone1998,Dukelsky1999,Volya2001}. Those which apply to the general case, such as the Richardson solution \cite{Richardson1963}, can only be solved numerically in a limited number of cases. Very recent results obtained by Ho et al.~\cite{Ho2010} obtain the energy gap of the BCS Hamiltonian without the need for the BCS approximation. However these results aren't as accurate for eigenvectors in the weak interaction regime. This means that to get an accurate result, direct diagonalisation is often still necessary. However, it is resource intensive on a classical computer to diagonalise the exact Hamiltonians for more than a few tens of qubits. Therefore it is still difficult to characterise systems for cases where $V_{ml}$ can't be held constant. Wu et al.~\cite{Wu2002} have shown that on a quantum computer it is possible to conduct a simulation of the BCS Hamiltonian efficiently, then use this simulation to extract information about the energy gap. While they concentrate on a nearest neighbour case, they also describe a case with general $V_{ml}$. Quantum simulation should thus be able to confirm whether the BCS Hamiltonian is correct for the metallic superconductors not covered by the assumption $V_{ml} = -V$. 

To conduct a simulation of the BCS Hamiltonian on a quantum computer, we need to map equation (\ref{BCSpair}) to the qubit Hamiltonian. This mapping is worked out by Wu et al.~\cite{Wu2002} and results in the BCS Hamiltonian taking the form:
\begin{equation}
H_{\mathrm{pair}} = \sum_{m=1}^{N} \frac{\varepsilon_{m}}{2}\sigma_{zm} + \sum_{m<l}^{N} \frac{V_{ml}}{2} \left( \sigma_{xm}\sigma_{xl} + r\sigma_{ym}\sigma_{yl} \right)
\label{bcsqubit}
\end{equation}
where $\sigma_{xk}$/ $\sigma_{yk}$ are the Pauli $X$/$Y$ on the $k^{\mathrm{th}}$ qubit, and $\varepsilon_{m}=\epsilon_{m} + V_{mm}$ and $r$ is a parameter determined by the mapping. 

As we are interested in determining the energy gap between the ground and first excited state, we start by preparing our system in a superposition of these two states.  This initial state is prepared using the technique proposed by Wu et al.~for preparing an initial state on an NMR computer \cite{Wu2002}. We first prepare our system of $N$ qubits in a state of all zeros, this is a basic ground state. We then apply spin flips so we have a basis state containing $n$ ones, where $n$ is the number of excitations in the BCS Hamiltonian. This is different from $N$ which is the number of modes in the BCS Hamiltonian and the number of qubits in our quantum computer. Using a quasi-adiabatic method of applying our Hamiltonian we prepare a superposition of the ground and first excited state, we discuss this step in more detail in section \ref{initialisation}. This generates an initial state which will allow us to extract data about the low-lying energy spectrum of our Hamiltonian, and thus determine the energy gap. 

To perform the time evolution and extract the data we use the phase estimation algorithm \cite{Cleve1998}, this involves performing the BCS Hamiltonian controlled on a set of ancilla qubits for a set of time intervals given by $2^{k-1}t$ for the $k^{\mathrm{th}}$ ancilla qubit. This encodes the data about the time evolution of the BCS Hamiltonian into the ancilla qubits. A quantum Fourier transform is then used to extract data from the system. In section \ref{evolution} we looked at the number of operations required to implement our unitaries in a controlled fashion, and in section \ref{extraction} we consider the number of operations needed to perform our quantum Fourier transform.  

We will consider conducting our simulation on an ancilla-driven quantum computer, where an auxiliary system is used to generate interactions in the main processing unit. The main advantage of such a system is that it allows us to generate entanglement between distant qubits without the need for swap gates. In terms of the BCS Hamiltonian this facilitates the simulation of  long range interactions, something difficult to do classically. While our results are applicable to ancilla driven qubit computers in general we will concentrate in particular on the  qubus quantum computer. The qubus is  a continuous variable ancilla that is used to generate interactions between the qubits \cite{Loock2008,Spiller2006}. The continuous variable bus is advantageous because detecting and controlling single photons is experimentally hard, while the continuous variable is easier to control. The qubus system generates gates between qubits deterministically without the need for measurement, but does require us to be able to generate an interaction between the field and a matter qubit \cite{Spiller2006}. 

In this paper we describe a method for simulating the time evolution of the BCS Hamiltonian then extracting the energy gap, using a similar technique to the one proposed by Wu et al.~\cite{Wu2002}. We show that using a qubus quantum computer gives significant advantages over an NMR quantum computer, and go on to show how we can obtain significant improvements over a na\"ive qubus implementation. This allows us to determine information about the ground and the first excited states of the Hamiltonian more efficiently than Wu et al.~\cite{Wu2002}. The simulation is limited in accuracy only by the number of terms used in the Trotter approximation to combine the non-commuting parts of the Hamiltonian, and the number of ancilla qubits. In both cases a linear increase in precision costs a linear increase in resources \cite{Brown2006}. Other than this, no approximations are used and we can otain results in the completely general case where the coupling between pairs can take any value for any pair. We can therefore access regimes unobtainable using classical approximation methods.

The paper is arranged as follows. In section \ref{qubus} we introduce the qubus quantum computer and discuss how we go about reducing the number of interactions with the bus needed to perform certain operations. Section \ref{interaction} contains three methods for producing the necessary unitaries for the BCS Hamiltonian on the qubus; subsection \ref{general} concentrates on the general case; subsection \ref{limited} looks at a case where it is possible to get an order of $N$ saving in return for a reduction of generality; and subsection \ref{nn} looks at fixed range interactions. Section \ref{total} puts the results from sections \ref{initialisation}, \ref{evolution}, and \ref{extraction} together to work out how many operations are needed as part of the longest single run of the qubus computer. We summarise our results in section \ref{conclusions}.

\section{The Qubus Quantum Computer} \label{qubus}
A qubus quantum computer is a hybrid system consisting of a processing unit made of qubits and a continuous variable field `bus' which generates interactions and transfers information between the qubits. Using the bus to generate interactions between distant qubits removes the need to either change calibration settings every time a new qubit is added or use swap operations to move qubits next to each other. There are several proposals for physical architectures which could potentially be used to build a qubus system. These include optical quantum systems \cite{Loock2008} and super-conducting systems \cite{Rodrigues2008}.

Interactions in the qubus architecture take the form of displacement operators applied to the continuous variable field. These can be written in the form \cite{Spiller2006}
\begin{equation} \label{int}
 D(\beta \sigma_{zk}) = \exp(\beta \sigma_{zk} a^\dagger - \beta^* \sigma_{zk} a) 
 \end{equation}
where $a^{\dagger}$, $a$ are the creation and annihilation operators, $\sigma_{zk}$ is the Pauli Z operator acting on the $k^{\mathrm{th}}$ qubit, $\beta = \chi t e^{i(\phi - \frac{\pi}{2})}$, and $\chi$ is the strength of the nonlinearity being used. In the qubus system we use $\phi=0$, corresponding to the position quadrature or $\phi = \frac{\pi}{2}$, corresponding to the momentum quadrature. 

\begin{figure}[ht]
\centering
     \includegraphics[width=135mm]{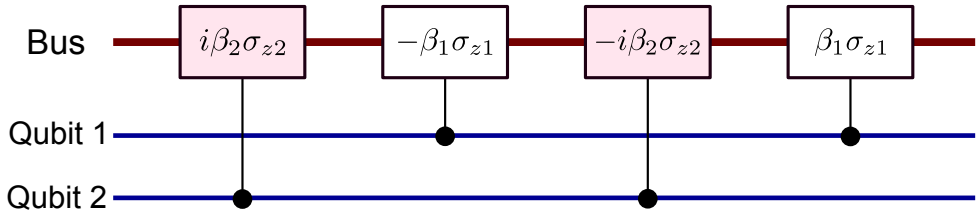} 
     \caption{A circuit diagram for implementing the C-Phase gate on the qubus. The boxes show displacements performed on the continuous variable field controlled by the qubits as in equation (\ref{int}). Shaded boxes represent an operation acting on the position quadrature of the bus, while unshaded boxes represent operations on the momentum quadrature.} \label{cphase}
\end{figure}

The displacement operators entangle the bus with the qubits. A qubit is entangled to either the position or momentum quadrature of the bus, and since these displacements are in orthogonal directions it is possible to create a maximally entangling gate. Using two qubits and one bus it is possible to generate a deterministic C-Phase gate by performing just four displacement operations on the bus. This scheme is illustrated in figure \ref{cphase} and is given by the operator $D(i\beta_{2}\sigma_{z2})D(\beta_{1}\sigma_{z1})D(-i\beta_{2}\sigma_{z2})D(-\beta_{1}\sigma_{z1})$, see \cite{Spiller2006}.

These operations are combined using the formula 
\begin{equation}
D(a)D(b) = \exp \left( \frac{ab^{*} - a^{*}b}{2} \right)D(a+b)
\end{equation}
where $a$ and $b$ are Pauli operators on the qubits and must commute (e.g.~because they act on different qubits). This sequence thus results in the interaction $\exp\left(2i\beta_{1}\beta_{2}\sigma_{z1}\sigma_{z2} \right)$, which when $2\beta_{1}\beta_{2} = \pi/4$ is equivalent under local unitaries to a C-Phase gate. 

\begin{figure}[ht]
\centering
     \includegraphics[width=135mm]{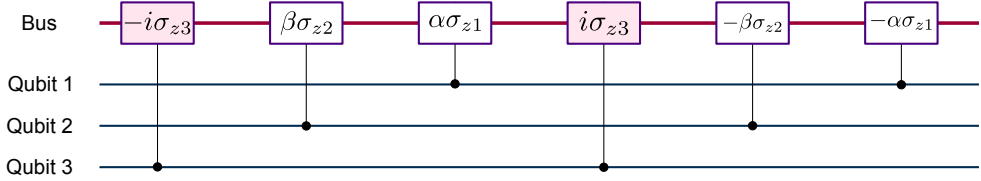}
     \caption{A reduced gate sequence for $\exp\left(i\alpha\sigma_{z1}\sigma_{z3} + i\beta\sigma_{z2}\sigma_{z3} \right)$ using 6 operations instead of 8. The boxes show displacements performed on the continuous variable field controlled by the qubits. Shaded boxes represent an operation acting on the position quadrature of the bus, while unshaded boxes represent operations on the momentum quadrature.} \label{Example}
\end{figure}

We will define the `na\"ive' method of implementing an operation on a qubus to be implementing a Hamiltonian by performing each C-Phase gate individually, disentangling the bus from the system between each set of gates. This technique requires four operations for each C-Phase gate within the desired interaction. If we take an interaction of the form $\exp \left(i\alpha\sigma_{z1}\sigma_{z3} + i\beta\sigma_{z2}\sigma_{z3} \right)$, then using the na\"ive method would require a total of eight operations to implement it. However, the term $\sigma_{z3}$ appears twice, therefore it is possible to reduce the total number of bus operations by rearranging the gate sequence, so the 3$^{\mathrm{rd}}$ qubit only needs to be interact with the bus twice. The gate sequence required to do this is shown in figure \ref{Example}, it uses only six operations. 

Provided there are `overlapping' terms such as the $\sigma_{z3}$ in the above example, it is possible to get a reduction in the number of operations required over the na\"ive method. The extent of this reduction depends upon the number of terms that overlap, how often these overlapping terms appear, and the level of generality required. We will now apply these techniques to reduce the number of operations needed to simulate the BCS Hamiltonian. 

\section{Implementing the Hamiltonian} \label{interaction}
We will now consider how to perform the BCS Hamiltonian on our qubus quantum computer, this stage is essential for both the initialisation process, where it is used for a semi-adiabatic evolution, and the time evolution stage.  The time evolution of the BCS Hamiltonian in terms of Pauli operators is shown in equation (\ref{bcsqubit}), which consists of three non-commuting Hamiltonians so that $H_{\mathrm{pair}} =H_{0} + H_{xx} + H_{yy}$

Taking the corresponding unitaries to these Hamiltonians we get
\begin{eqnarray}
& U_{0} &=  \exp \left(i \sum_{m}^{N} \frac{\varepsilon}{2} \sigma_{zm} \right)  \label{H01}\\
&  U_{xx} &= \exp \left( i\sum_{m<l}^{N} \frac{V_{ml}}{2} \left( \sigma_{xm}\sigma_{xl} \right) \right) \label{Hxx1} \\
&  U_{yy} &= \exp \left( i\sum_{m<l}^{N} \frac{V_{ml}}{2}r \left( \sigma_{ym}\sigma_{yl} \right) \right)\,. \label{Hyy1}
\end{eqnarray}
Since these are non commuting unitaries, $\exp(H_{\mathrm{pair}}) \neq \exp(H_{0})\exp(H_{xx})\exp(H_{yy})$. To get around this we use the Trotter approximation \cite{Trotter1959,Suzuki1993} splitting the time $t$ into small segments $\tau$. In the first order case this is given by $\exp(H_{\mathrm{pair}}) \approx \exp(H_{0}\tau)\exp(H_{xx}\tau)\exp(H_{yy}\tau) + \mathrm{O}(\tau^2)$, and in the second order case it is given by  $\exp(H_{\mathrm{pair}}) \approx \exp(H_{0}\tau/2)\exp(H_{xx}\tau/2)\exp(H_{yy}\tau)\exp(H_{xx}\tau/2)\exp(H_{0}\tau/2) + \mathrm{O}(\tau^3)$. Longer time intervals are built up by performing the short time evolution multiple times. 

In equation (\ref{H01}), $U_0$ is a sum of local operations and can be implemented using N local unitaries. As we are considering a situation where we have individual addressability of the qubits, this is straight forward and we won't discuss it further.  In equations (\ref{Hxx1}) and (\ref{Hyy1}), $U_{xx}$ and $U_{yy}$ are equivalent under local unitaries to 
\begin{equation}
U_{zz} = \exp \left(i \sum_{m<l}^{N} \frac{V_{ml}}{2} \left( \sigma_{zm}\sigma_{zl} \right) \right)\,. \label{Hzz1}
\end{equation}
We now consider how to implement equation (\ref{Hzz1}) in three different cases.
\subsection{The fully generalised case}\label{general}
Equation (\ref{Hzz1}) is a sum of $(N^2 - N)/2$ terms, each requiring one gate of the form $\exp(i\beta_{m}\beta_{l}\sigma_{zm}\sigma_{zl})$. For the simplest possible implementation on a qubus computer, we perform each of these gates individually requiring 4 bus operations per gate. This means $U_{zz}$ requires a total of $2N^2-2N$ bus operations to perform. 

This technique uses the simplest possible procedure for generating the necessary interactions, and it should be possible to get reductions by reusing our bus. However, if we wish to be able to set each of our constants, $V_{ml}$, separately then we have significant constraints on how we can do this. A simple lower bound on the number of bus operations needed can be found by considering the number of individual constants. For a system being modelled on $N$ qubits, we require $(N^2 - N)/2$ constants, which have no dependence upon each other. Since our operations on the bus occur in pairs, and each pair can only produce one constant we can see that it would be impossible to generate this using less than $N^2 - N$ bus operations. 

We now consider how we would go about generating the necessary bus operations. We want qubit 1 to interact with all the other $N-1$ qubits, with completely independent constants. The simplest way to do this would be to connect qubit 1 to one quadrature of the bus, then connect the other $N-1$ qubits to the other quadrature of the bus. Now, when we consider qubit 2, we can see that we have already generated the interaction between this and qubit 1, therefore it only needs to interact with the remaining $N-2$ qubits. However we want the constants generated from these interactions to be completely independent from the ones generated by interacting qubit 1 with the others. If we hold too many qubits on the bus, this independence isn't possible, therefore after each step we need to disconnect all of the qubits from both quadratures of the bus. 

\begin{figure}[ht]
\centering
     \includegraphics[width=135mm]{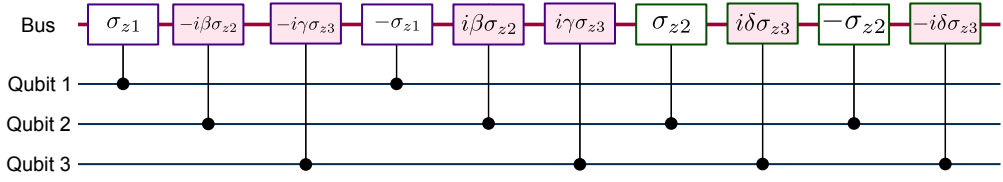}
     \caption{A circuit for simulating $U_{zz}$ for 3 qubits with generalised constants $V_{ml}$ in $10$ operations. The boxes show displacements performed on the continuous variable field controlled by the qubits. Shaded boxes represent an operation acting on the position quadrature of the bus, while unshaded boxes represent operations on the momentum quadrature.} \label{Z3}
\end{figure}  

We therefore consider an $N-1$ step process, where for the $a^{\mathrm{th}}$ step we connect qubit $a$ to one quadrature of the bus, and the other $a-1$ qubits to the other quadrature before disconnecting qubit $a$ then all the others. The total number of bus operations needed for this is given by
\begin{equation}
\sum^{N}_{a=2} 2a = 2\left( \sum_{a=1}^{N}a \right) - 2 = N^2 + N -2\,.
\end{equation}
A circuit for generating interactions between 3 qubits using this technique is shown in figure \ref{Z3}. 

\begin{figure}[ht]
\centering
     \includegraphics[width=135mm]{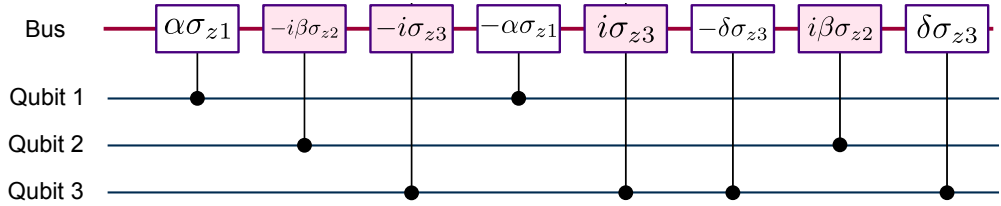}
     \caption{A circuit for simulating $U_{zz}$ for 3 qubits with generalised constants $V_{ml}$ in $8$ operations. The boxes show displacements performed on the continuous variable field controlled by the qubits. Shaded boxes represent an operation acting on the position quadrature of the bus, while unshaded boxes represent operations on the momentum quadrature.} \label{smallerZ3}
\end{figure}  

One way of reducing the number of operations further is to leave some qubits on the bus at the end of each step. We now consider the case where we first attach qubit 1 to the first quadrature of the bus and the other $N-1$ qubits to the second quadrature, as before. However when we come to disentangle the qubits, we leave qubit 2 entangled to the second quadrature of the bus. Step two then involves attaching qubits 3 through to $N$ to the first quadrature of the bus thus generating entanglement between all these qubits and qubit 2. In the disentangling step all the qubits are removed from the bus except qubit 3, which is used for step three. In this scenario qubit 1 and 2 interact with the bus twice (one connect and one disconnect) each, qubit 3 interacts with the bus four times (two connects and two disconnects) etc. This occurs all the way up to qubit $N$ which interacts with the bus $2(N-1)$ times. This leads to a total number of operations given by 
\begin{equation}
2\left( \sum_{a=1}^{N-1}a \right) + 2 = N^2 - N + 2\,.
\end{equation}
This is $2N-4$ operations less than our previous result and very close to our lower bound. A circuit for performing this technique using 3 qubits is shown in figure \ref{smallerZ3}. It is less obvious that in this case the constants can be completely general as one constant from each step will carry through to the next step. However, since all the constants in the next step can be set arbitrarily it is possible to achieve this. If there is no interaction between two qubits, e.g.~between qubits 2 and 3, then the higher numbered qubit is skipped in that step and added in during a later step. This will never require extra operations and will in some cases provide a saving. 

\subsection{A limited case}\label{limited}
While the completely general case is important, many physical systems do not require this level of generality. If we are prepared to restrict our constants so that they become dependent upon each other then it is possible to get savings of $O(N)$ over our previous techniques. This provides a saving for a a range of different scenarios including ones where the strength of the interaction drops off exponentially with distance. We will begin by considering the simplest possible case of wanting to create the $U_{xx}$ and $U_{yy}$ that are local unitaries away from $U_{zz}$ in equation (\ref{Hzz1}) when $V$ has no dependence upon either $m$ or $l$. The minimal number of bus operations needed to generate this can be worked out by noting that for the sum given, both $m$ and $l$ can take $N-1$ possible different values. This means that $2(N-1)$ bus operations are needed to cover all the possible values of $m$ and similarly for $l$. This results in a total number of bus operations given by $4(N-1)$. 

\begin{figure}[ht]
\centering
   \includegraphics[width=135mm]{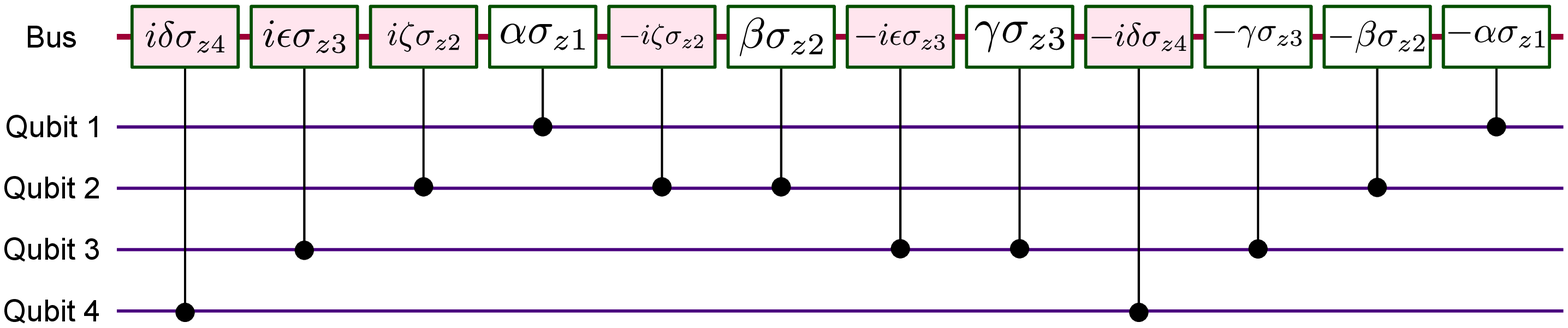}
   \caption{A circuit for simulating $U_{zz}$ with a strong dependence between the constants. The boxes show displacements performed on the continuous variable field controlled by the qubits. Shaded boxes represent an operation acting on the position quadrature of the bus, while unshaded boxes represent operations on the momentum quadrature.} \label{Reduced}
 \end{figure}

A method of performing the operations that meets this bound involves attaching qubits 2 through to $N$ to the first quadrature of the bus. Qubit $1$ is then attached to the second quadrature, generating entanglement between it and all the other qubits. Thus far, this matches the method in section \ref{general}. However now, instead of removing all the qubits from the bus, we disconnect qubit 2 from the first quadrature of the bus and then reattach it to the second quadrature. This generates entanglement between it and qubits 3 through to $N$. We continue this process, removing qubits 2 through to $N-1$ from the first quadrature, then attaching them to the second quadrature. Finally when we come to remove qubit $N$ from the bus, we don't reattach it to the second quadrature. This leaves the first quadrature of the bus completely empty and $N-1$ qubits entangled to the second quadrature of the bus. The final step involves removing all remaining qubits from the second quadrature of the bus. This means every qubit interacts with the bus 4 times (2 connections and 2 disconnections), apart from the first and last qubit, which only interact with the bus twice. The total number of bus operations needed is given by $4N - 4$, therefore this technique meets the lower bound exactly. A circuit showing the sequence of bus operations for this technique in the four qubit case is shown in figure \ref{Reduced}. 

This technique can give constants more general than a fixed $V$. The constants attached to each qubit give a matrix of values for $V$ that are dependent upon $m$ and $l$ and are linked as shown below
\begin{equation*}
\begin{array}{cccc}
& 2 & 3 & 4 \\
1 & \alpha\zeta & \alpha \epsilon & \alpha \delta \\
2 & &\beta \epsilon & \beta \delta \\
3 & & & \gamma \delta\,.  
\end{array} \label{matrix}
\end{equation*}
The entries of the matrix above represent the strength of the interaction between the numbered qubits given the constants in figure \ref{Reduced}, so for example the second column of the first row represents the strength of the interaction between qubit 1 and qubit 3. By looking at the matrix we can see that it wouldn't be possible to set values for $\alpha, \beta, \gamma, \delta, \epsilon$ and $\zeta$ such that all six entries of the matrix could be set completely arbitrarily. We can see this easily because $\alpha\epsilon / \alpha\delta$ has to equal $\beta \epsilon / \beta\delta $ if $\alpha, \beta, \delta$ and $\epsilon$ are complex constants; however if we set all 6 entries of our matrix randomly then this isn't always the case.  

This dependence allows numerous cases, including a exponential fall off with distance or other more complicated dependencies. An example of exponential decay would be setting $\alpha=1$, $\beta=2$, $\gamma=3$, $\delta=1/4$, $\epsilon=1/2$ and $\zeta=1$. The possible constants provide an interesting set of cases relevant to realistic situations. 

\subsection{Fixed range interactions}\label{nn}
One case not covered by our limited case, in which it is obviously possible to get improvements, is nearest neighbour interactions. This would consist of a chain of qubits and is explored by Louis et al.~\cite{Louis2007} in the context of building cluster states. This involves a $U_{zz}$ of the form 
\begin{equation}
U_{zz} = \exp \left( \sum_{m=1}^{N-1} \frac{V_{m,m+1}}{2} \left( \sigma_{zm}\sigma_{z(m+1)} \right) \right)\,. \label{Hzz2}
\end{equation}
Louis et al.~find that for a chain of $N$ qubits, it is possible to generate nearest neighbour interactions using just $2N$ bus operations. This can be expanded to include a $U_{zz}$ with both nearest neighbour and next-to-nearest neighbour interactions,  a scenario which would require $4N - 4$ bus operations. 

It is possible to derive a general result for fixed range interactions by introducing a new variable $p$, where $p$ represents the range of the interaction. When $p =1$ we consider only nearest neighbour interactions, when $p= 2$ we consider nearest neighbour and next-to-nearest neighbour, etc.. We split our terms into a cycle of several steps, where we leave one qubit active with the bus at the end of each step, this allows us to keep the interaction strengths between various qubits completely generalised. For the first step, $p+1$ qubits need to interact with the bus twice, this represents qubit 1 and the $p$ qubits connected to it. The next set of steps need $p$ qubits to interact with the bus twice, since we assume the $1^{\mathrm{st}}$ qubit in each step is already connected to the bus we only need to attach its $p$ nearest neighbours. We require $2p$ qubits operation on the bus for every single step except for the very first one and the $p$ final steps. We therefore need to perform $2p$ operations a total of $N - p - 1$ times. Finally for the last set of steps there are no longer $p$ qubits left in the chain connected to our active qubit, this means the number of bus operations required reduces from $2(p-1)$ to $0$ in intervals of $2$. Therefore, for an interaction length $p$, the number of operations on the bus needed to generate $U_{zz}$ is given by 
\begin{equation}
2(p+1) + 2p(N-p-1) + 2\sum_{a=1}^{p} (p-a) = 2pN - p^2 - p +2\,. \label{pneighbour}
\end{equation}

We can check that this equation agrees with our previous results. If we take the nearest neighbour case, then $p=1$ and we find that the total number of operationss needed to generate $U_{zz}$ is $4N$. If we take the case where all the qubits interact with each other, $p=N-1$ therefore $N^2 - N +2$ bus operations are needed to generate $U_{zz}$. This demonstrates a technique that allows the generation of interactions with a cut off at an arbitrary distance. It should be possible to obtain further savings in these cases using a technique similar to that described in section \ref{limited}.

\section{Initialisation}\label{initialisation} 
To initialise our system we need to generate a superposition of the ground and first excited state for our BCS Hamiltonian. To do this, we can use the same method as Wu et al.~\cite{Wu2002} for the initialisation procedure within an NMR system. We begin by preparing our system so that we have $N$ qubits all in state $|0\rangle$, where $N$ is the number of modes in the BCS Hamiltonian, this is a basic ground state. By applying spin flips it is possible to transform this state so that $n$ qubits are in state $|1\rangle$, where $n$ is the number of excitations within our BCS Hamiltonian. This generates a computational basis state and the process is trivial to perform on a qubus quantum computer. 

To get our qubus quantum computer to be in a combination of the ground and first excited state of the BCS Hamiltonian we need to implement an adiabatic form of our time evolution given by
\begin{equation}
U_{ad}(k\tau) \approx e^{-H(k\tau)\tau} \cdots e^{-iH(2\tau)\tau}e^{-H(\tau)\tau} + O(\tau^2)
\end{equation}
where $\exp[-iH(j\tau)\tau] \approx \exp(iH_{0}\tau)\exp[-ic(j\tau)H_{1}\tau]$, $j = 1, \dots, S$, $S\tau = T$, and $c$ is a function which varies slowly with time, $t$, that goes from $0$ at $t=0$ to $1$ at $t=T$. In this case $H_{1} = H_{xx} + H_{yy}$. We are gradually increasing the strength of the interactions within the Hamiltonian so that we go from a basis state of $H_{0}$ to a state that is described by the BCS Hamiltonian. If we conducted this evolution in accordance with the adiabatic condition ($S\gg\pi/(\tau\Delta)$ where $2\Delta$ is the energy gap between the ground and first excited state), then we would end up in the ground state of the BCS Hamiltonian. However, instead we relax the adiabatic condition and as a result end up with a component of the first excited state mixed in with the ground state. In the case of the BCS Hamiltonian the short time approximation is valid when $\tau \ll 1/d$ where $d$ is the level spacing, and $\tau$ the length of a single time interval. This gives $S\gg\pi d/{\Delta}$. Brown et al.~\cite{Brown2006} add in another factor which represents the precision of the desired energy gap, and is given by $\delta$. This results in the condition $S = \pi d  / \delta \Delta$. Taking $d/\Delta=0.1$ from Wu et al.~\cite{Wu2002} then the total number of time steps required is given by 
\begin{equation}
S = \pi/\delta \,. \label{S}
\end{equation}

The time evolution here can be implemented using the method described in section \ref{interaction}. As we are considering a short evolution we can consider using only the first order Trotter approximation, similar to Wu et al.~\cite{Wu2002}. This means $H_{0}$ can be implemented using single local unitaries on each qubit and $H_{1}$ will require twice the number of operations needed to perform $H_{zz}$ plus $4N$ local unitaries to transform our $U_{zz}$ terms to $U_{xx}$ and $U_{yy}$. We can therefore see that the number of operations per time step, $I(N)$ will be: 
\begin{equation}
I_{G}(N)=2N^2 + 3N +4 \label{IG}
\end{equation}
in the general case, $I(N)=13N - 8$ in the limited case and 
\begin{equation}
I_{L}(N)=4pN + 5N - 2p^2 -2p +4 \label{IL}
\end{equation}
in the case of fixed range interactions. To work out the total number of operations needed for the initialisation procedure we multiply $I(N)$ by $S$. 

\section{The Phase Estimation algorithm}\label{evolution}
Once we have created the superposition of the ground and excited state we want to use the phase estimation algorithm for our data extraction procedure. The phase estimation algorithm has two stages, the first stages implements the unitary representing the BCS Hamiltonian and therefore encodes the data about our evolution into a series of ancilla qubit, the second uses a quantum Fourier transform to extract the data. 

\begin{figure}[ht]
\centering
\includegraphics[width=12cm]{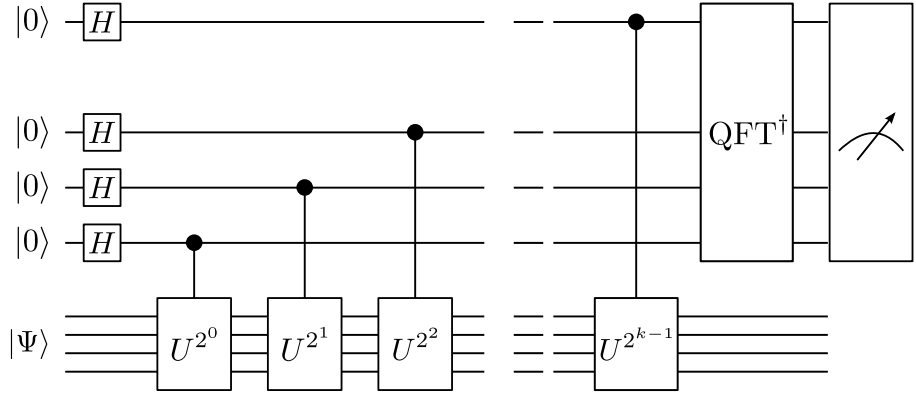}
\caption{A circuit to perform the phase estimation algorithm, the sequence of controlled unitaries encode phase information from the input state $|\Psi\rangle$ in the ancilla which is then extracted using the QFT.} \label{PE}
\end{figure}

A circuit for performing the phase estimation algorithm is shown in figure \ref{PE}. To illustrate the workings of the phase estimation algorithm we will consider our first register to be in a single eigenstate,  $|u\rangle$, of $U$, therefore implementing $U$ will leave $|u\rangle$. Implementing $CU^{2^{j}}$, i.e.~controlled on an ancilla qubit results in 
\begin{equation}
CU^{2j}\frac{1}{\sqrt{2}}\left ( |0\rangle + |1 \rangle \right ) |u \rangle = \frac{1}{\sqrt(2)}\left ( |0\rangle + e^{i2^{j}\Phi}|1\rangle \right) |u\rangle \,.
\end{equation}
where $e^{i\Phi}$ given by $U|u\rangle = e^{i\Phi}|u\rangle$ is the eigenvalue of $|u\rangle$. We now consider applying these gates from $j=0$ to $j=k-1$ using $k$ ancilla qubits, with the net result 
\begin{equation}
\fl \frac{1}{\sqrt{2^k}} \left ( |0\rangle + e^{i2^{k-1}\Phi}|1\rangle \right ) \dots \left (|0\rangle + e^{i2\Phi} |1\rangle \right ) \left( |0\rangle + e^{i\Phi}|1\rangle \right)|u\rangle = \frac{1}{\sqrt{2^k}} \sum^{2^{k-1}}_{y=0}e^{i\Phi y}|y\rangle |u\rangle
\end{equation}
Tracing out the register containing the eigenstate leaves us with our $k$ ancilla qubits in the state $\sum^{2^k-1}_{y=0}e^{i\Phi y}|y\rangle$. Applying the inverse quantum Fourier transform encodes the phase into the register of ancilla qubits \cite{Nielsen2000a}. 

Assuming our register starts in an exact eignestate the final result of the phase estimation procedure is $|\tilde{\Phi} \rangle |u\rangle$, where $|\tilde{\Phi} \rangle$ is a k qubit approximation of $|\Phi \rangle$. If the first register starts in an approximation of the eigenstate, then the phase estimation algorithm results in the second register being transformed into $|\Phi \rangle$ with a probability that increases with the accuracy of the eigenstate and with $k$ \cite{Nielsen2000a}. Similarly if the register starts in a superposition of eigenstates, then the ancilla register ends up being in one of the resultant eigenvalues with a probability that depends on how much the corresponding eigenstate contributes to the initial input state. In our case, since we start in a superposition of the ground and the first excited state we can extract the eigenvalue for the ground state, and the eigenvalue for the first excited state by using several runs of the computation. This will allow us to work out the energy gap. 

\begin{figure}[ht]
\centering
\includegraphics{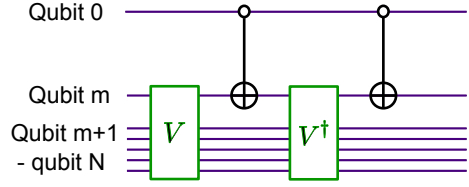}
\caption{A circuit to make a unitary $U$ controlled on an ancilla qubit. In this case $U=V^2$, and $U$ consists of pairs of Pauli $Z$ operations where each pair of operations acts on qubit $m$.} \label{Control}
\end{figure}

To implement our unitary as part of the phase estimation algorithm involves performing our operations in a controlled fashion, dependent upon the state of an ancilla qubit.  In section \ref{interaction} we discussed various techniques for performing our unitary, $U_{zz}$. We note that it is possible to make our $U_{zz}$ controlled using a very simple control sequence such as the one shown in figure \ref{Control}, provided each pair of operations acts on one common qubit. Since we need to perform CNOT operations after our sequence we can see the technique outlined in section \ref{general} where we don't leave qubits connected to the bus, is ideal for use in this scenario. 

Previous work on the qubus has outlined how to use the architecture to generate CNOT gates \cite{Spiller2006}, each gate requires 4 bus operations and 2 local unitaries. We also note that it is possible to transform our operations $U_{zz}$ to the form $U_{xx}$ or $U_{yy}$ using two local operations per qubit. Therefore we can work out the total number of operations needed to generate our two qubit interactions in a controlled fashion. In section \ref{general} we split generating our $U_{zz}$ into cycles where we interacted each qubit with all qubits numbered higher than it. We now want to consider how many operations would be required to implement these cycles in a controlled fashion. A cycle which interacts qubit $m$ with all the qubits numbered higher than $m$ required $2(N-m+1)$ operations to generate. To implement this in a controlled fashion we use a control procedure such as the one shown in figure \ref{Control}, therefore the total number of operations required per cycle is $4(N-m+4)$ including the gates required for our CNOT operations. To implement $U_{zz}$ we will require $2(N^2+7N-8)$ operations. To transform $U_{zz}$ to $U_{xx}$ or $U_{yy}$ requires an additional 2 operations per qubit bringing the total to $2(N^2+8N-8)$. 

\begin{figure}[ht]
\centering
\includegraphics[width=10cm]{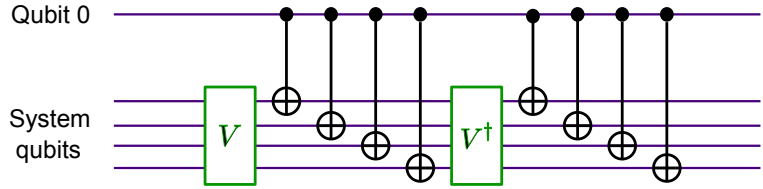}
\caption{A sequence to make $N$ local unitaries controlled on an ancilla qubit, where $V^2 = U_{L1} \otimes \dots \otimes U_{LN}$.} \label{Single}
\end{figure}

The most efficient technique for making our local unitaries controlled involves performing our CNOT operations in a sequence, as in the circuit in figure \ref{Single}, as a result a sequence which requires a single local unitary on each qubit will need $8N+4$ operations.

When combining our unitaries $U_{xx}$, $U_{yy}$ and $U_{0}$, we use the Trotter approximation. Since we need a more accurate $U$ than for initialisation we will use the second order Trotter approximation \cite{Trotter1959,Suzuki1993} which is given by
\begin{equation}
U_{BCS} \approx [U_{0}(\tau/2)U_{xx}(\tau/2)U_{yy}(\tau)U_{xx}(\tau/2)U_{0}(\tau/2)] \,. \label{Eq:DE:2trotter}
\end{equation}
where $\tau$ is the length of a single time interval. Figure \ref{PE} shows a circuit for performing the phase estimation algorithm. We will choose $U$ to be an implementation of the BCS Hamiltonian for a single time interval in the Trotter approximation, $U^2$ will require two time intervals, $U^4$ will need four intervals etc. This will allow us to work out the number of operations required as a function of the number of ancilla qubits. We therefore find that the total number of operations required for running the local unitaries in the phase estimation algorithm with $k$ ancillas is given by $(2^{k}-1)$ times the number of operations needed to implement $U$, therefore we have
\begin{equation}
P_{G}(N) = (2^{k}-1)(6N^2 + 64N -40) \label{PG}
\end{equation}
in the completely general case, and 
\begin{equation}
P_{L}(N) = (2^{k}-1)(12Np - 6p^2 - 6p + 70N -40) \label{PL}
\end{equation}
in the case of limited range interactions. The number of ancillas required is chosen dependent upon the desired accuracy. 

\section{Data extraction}\label{extraction}
As well as performing our controlled unitaries, the phase estimation algorithm requires us to implement the quantum Fourier transform (QFT), so we provide an efficient technique for performing the QFT on the qubus quantum computer. This provides us with significant savings over a na\"ive qubus implementation. 

\begin{figure}[ht]
\centering
   \includegraphics[width=155mm]{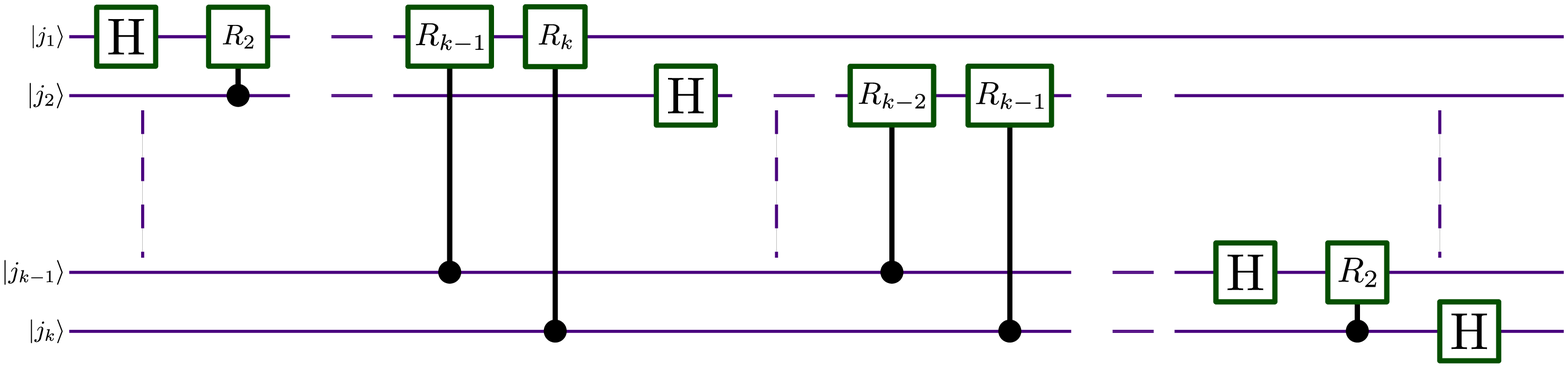}
   \caption{A circuit diagram for the quantum fourier transform on $k$ qubits, where $R_{a}$ are rotations on qubit $a$, and H is the Hadamard operation. The controlled unitaries which $R_{a}$ are part of are described by the $CR_{a}$ in equation  (\ref{Ra}).} \label{FT}
\end{figure}

The QFT involves performing controlled rotation operators on the qubits in a form shown in figure \ref{FT}, where the gate $CR_{a}$ corresponds to a controlled phase gate of the form
\begin{equation}
CR_{a} = \left ( 
\begin{array}{cccc}
1  & 0 & 0 & 0 \\ 0 & 1 & 0 & 0 \\ 0& 0 & 1 & 0 \\0& 0&  0 &  e^{2\pi i/2^a}
\end{array}
\right )\,. \label{Ra}
\end{equation}
For each desired $CR_{a}$ we perform an operation 
\begin{equation}
CR'_{a} = \left ( 
\begin{array}{cccc}
e^{\pi i/2^{a+1}} & 0 & 0 & 0 \\ 0 & e^{-\pi i/2^{a+1}} & 0 & 0 \\ 0& 0 & e^{-\pi i/2^{a+1}} & 0 \\0& 0&  0 &  e^{\pi i/2^{a+1}}
\end{array}
\right )\,.
\end{equation}
We then correct these gates using local unitaries on each qubit, of the form 
\begin{equation}
C_{a} = \left (
\begin{array}{cc}
e^{-\pi i/2^{a+2}} & 0 \\ 0 & e^{3\pi i/2^{a+2}}
\end{array}
\right )\,.
\end{equation}
We chose to use this method of performing operations because of the limitations of the qubus quantum computer. The gates $CR'_{a}$ are gates which are relatively easy to perform, and as we can apply general local unitaries the $C_{a}$ gates don't present a problem. 

Performing a QFT on $k$ qubits involves a total of $\frac{1}{2}(k^2 - k)$ controlled phase gates. In a na\"ive case where each phase gate requires 4 operations to perform, we would require $2(k^2 - k)$ interactions with the bus. We can consider possible ways to reduce the total number of operations needed using a similar technique to the one described in section \ref{limited}, since the size of the rotation we need to apply decreases exponentially with the separation of the qubits in the register. The principle differences between our technique here and our previous technique is the need to apply Hadamard gates between each set of controlled phase operations, and the local unitary corrections. As our corrections commute with the $CR'_{a}$, we don't need to perform these straight away. However, these corrections do not commute with the Hadamard, so the corrections for each individual qubit need to be performed before the Hadamard is performed on that qubit. Multiple corrections are performed simply by taking the product of the necessary corrections for every single controlled phase gate the qubit is part of, regardless of whether it acts as a target or control qubit. 

\begin{figure}[ht]
\centering
   \includegraphics[width=155mm]{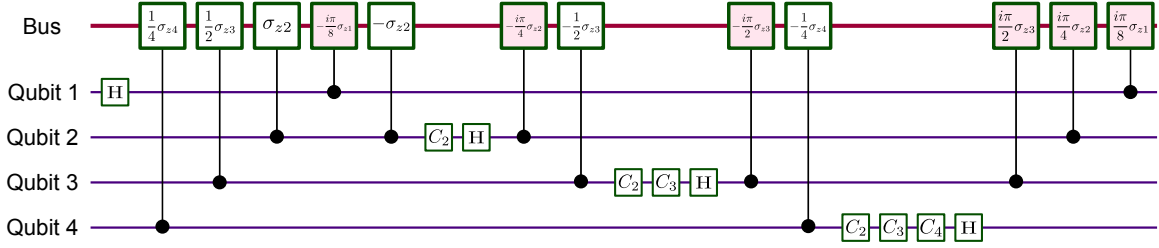}
   \caption{A diagram showing how to perform a 4 qubit Fourier Transform on the qubus quantum computer. The boxes show displacements performed on the continuous variable field controlled by the qubits or local operations. Shaded boxes represent an operation acting on the position quadrature of the bus, while unshaded boxes represent operations on the momentum quadrature.} \label{FTQB}
\end{figure}

To obtain the full QFT sequence we entangle qubit 1 with the first quadrature of the bus, then all the other $k-1$ qubits with the second quadrature. We then disentangle qubit 1 and qubit 2, before applying a correction and a Hadamard operation to qubit 2. Qubit 2 is then entangled to the first quadrature of the bus, thus interacting it with the other $k-2$ qubits. We repeat this process shifting successive qubits from the second quadrature of the bus to the first, applying a correction and a Hadamard operation in between the two operations. Once qubit k has been removed from the second quadrature of the bus, it doesn't need to be reconnected although the correction and the Hadamard operation need to be performed. We then remove qubits $2$ through to $k-1$ from the first quadrature of the bus and our QFT is complete. In this sequence qubits $1$ through to $k-1$ interact with the first quadrature of the bus twice (once to connect them, and once to remove them) and qubits $2$ through to $k$ interact with the second quadrature of the bus twice. This results in a total of $4k - 4$ bus operations. A diagram with suitable phase factors for performing the QFT in the 4 qubit case is shown in figure \ref{FTQB}. 

We then need to consider the corrections we need to apply to our qubits. In the worst case scenario we would need to apply a correction for every single controlled rotation a qubit is part of, meaning we'd need $k^2-k$ local unitaries. Since we are considering diagonal matrices, and have the ability to perform arbitrary unitaries, it is possible to reduce this down to $2k-2$ by combining the corrections performed on each qubit into those performed before the Hadamard and those performed after. This is feasible because the matrices are diagonal, so we can work out what correction we need to apply efficiently on a classical computer. In the case of the phase estimation algorithm we are measuring in the Z-basis straight after performing our QFT. This means it is possible to reduce the number of corrections further by simply not performing those which occur after the Hadamard. This results in a total of $k-1$ corrections being applied and therefore a total number operations for the QFT given by 
\begin{equation}
N_{FT} = 6k-5 \label{NFT}
\end{equation}
including the $k$ necessary Hadamard gates. 

\section{The total number of operations needed for a simulation}\label{total}
We now want to consider the total number of operations required for simulating the BCS Hamiltonian. The savings we showed in section \ref{interaction} are mainly applicable to the initialisation procedure, however even for a small number of ancilla qubits the number of operations required by the phase estimation algorithm dominates to such a degree that it is only worth considering two cases, the general case and the case of limited range interactions. In the general case we find that the total number of operations needed to implement our entire algorithm for a single run is given by 
\begin{equation}
T_{G} = P_{G}(N) +SI_{G}(N) +N_{FT}
\end{equation} from equations (\ref{S}), (\ref{IL}), (\ref{PG}) and (\ref{NFT}). This gives us
\begin{equation}
T_{\mathrm{G}} = (2^{k}-1)(6N^2 + 64N -40) + 6k-5 + \frac{0.1\pi}{\delta}(2N^2 + 3N +4) \,.
\end{equation} 
Using a similar argument for the limited range case we find, 
\begin{equation}
T_{\mathrm{L}} = (2^{k}-1)(12Np - 6p^2 - 6p + 70N -40) + 6k - 5 + \frac{0.1\pi}{\delta}(4pN + 5N - 2p^2 -2p +4) .\,
\end{equation}
To make a comparison with Wu et al.~\cite{Wu2002} we need to rewrite $k$ in terms of precision. The level of precision available can be expressed as a function of $2\pi$ such that the smallest phase difference we can detect is given by $2\pi/2^{k}$, to have an equivalent precision to Brown et al.~\cite{Brown2006} we want the smallest phase difference we can detect to be given by $\delta$, therefore $2^k = 2\pi/\delta$. The total number of operations is seen to be
\begin{equation}
T_{\mathrm{Gk}} \approx \frac{0.1\pi}{\delta} (122N^2 + 1283N - 796)
\end{equation}
in the general case, and
\begin{equation}
T_{\mathrm{Lk}} \approx \frac{0.1\pi}{\delta} (244Np - 122p^2 -122p + 1405N -796)
\end{equation}
in the limited range case. We can now compare our results to those found by Wu et al.~\cite{Wu2002} who claim to require significantly more than $3N^4$ operations, however this excludes the initialisation procedure. Since they use the first order Trotter approximation throughout we will derive an upper bound by using the same number of operations for the initialisation procedure as the longest run of the computer. We will incorporate the precision such that 
\begin{equation}
T_{\mathrm{NMR}} = \frac{6}{\delta}N^4
\end{equation}
in the nearest neighbour case to leading order. For our qubus system in the nearest-neighbour case we have 
\begin{equation}
T_{\mathrm{qubus}} = \frac{\pi}{\delta}(164.7N - 104) \,.
\end{equation}
We can therefore see that provided $N\ge 5$ our qubus system requires less operations than an equivalent NMR simulation. For $N<5$ it would be possible to get savings by using a similar data extraction procedure to NMR, running the simulation for several time intervals and working out the probability of a single qubit being in $|1\rangle$. While this would require more runs and have a poor scaling it is suitable for small systems. In the case of $N=10$ our qubus system requires $785,430 \approx 8\times 10^5$ operations, while the NMR system requires $6\times10^6$ operations, therefore we can already see a significant difference. Using a similar number ($\approx 6\times10^6$) of operations on our qubus system it would be possible to generate operations for a BCS Hamiltonian of 72 qubits in the nearest-neighbour case, and 26 qubits in the general case. 

\section{Conclusions}\label{conclusions}
In this paper we have discussed how to simulate the BCS Hamiltonian on a qubus quantum computer. Our results are also applicable to general ancilla-based schemes. We show that using a na\"ive method of performing the time evolution, we require $O(N^2)$ operations, which is an $O(N^3)$ saving over a simulation on an NMR computer \cite{Wu2002,Brown2006}. It is possible to get further reductions in the number of gates needed in the general case, where we get almost a factor of 2 savings over our na\"ive method, and in certain special cases where $O(N)$ saving is possible. In the general case we need $3N^2 + 2N +6$ operations, including local unitaries, for each time step of the evolution. For the special cases, which include the case where interactions are decaying exponentially with distance, only $17N - 12$ operations are required for each stage in the time evolution. We show that the general case requires only 6 operations more than our lower bound, and that our special cases achieve their overall lower bounds. We also look at the nearest-neighbour case of the BCS Hamiltonian which only requires $2N$ operations for $U_{zz}$ \cite{Louis2007} so $11N$ operations to perform in the general case. Building upon this we look at interactions of length $p$, where $p=1$ is the nearest-neighbour case, $p=2$ nearest-neighbour and next to nearest-neighbour etc. For this case, we characterise the number of operations needed to perform each stage of the time evolution as $5N + 6pN - 3p^2 - 3p +6$. 

We then apply these efficiency savings to making our operations controlled so we can implement the phase estimation algorithm. In this case we find that we can't get the O$(N)$ savings over our na\"ive method. However, we can still obtain significant savings, and require O$(N^3)$ less operations than an NMR computer. As we are using the phase estimation algorithm we also demonstrate how to obtain efficiency savings when performing a QFT on the qubus quantum computer, reducing the number of operations required by $O(N)$ to $6N -5$ per QFT. This significant improvement will allow efficient extraction of the energy gap from our system using the phase estimation algorithm. The efficiency savings for the quantum Fourier transform are applicable to other quantum algorithms, such as Shor's algorithm. 

Wu et al.~\cite{Wu2002} give a minimum bound on the number of operations required to simulate a BCS Hamiltonian described by 10 qubits on an NMR quantum computer. If we consider being limited to a similar number of operations, we show that on the qubus quantum computer we can simulate a BCS Hamiltonian described by 72 qubits. Thus our qubus quantum computer has significantly better efficiency than the NMR computer, and hence we will be able to increase the size of the system we can simulate on an early quantum computer. 

We have shown that the qubus quantum computer has significant advantages in the number of elementary operations required compared to an NMR computer, this gives O$(N^3)$ savings when simulating the BCS Hamiltonian. As we have demonstrated an O$(N)$ improvement over the na\"ive case of performing the quantum Fourier transform we have every reason to believe that these savings will apply in a wide range of cases. In other work we have shown similar improvements for generating cluster states \cite{Horsman2010}, illustrating the generality of the techniques we have described here in the context of simulating the BCS Hamiltonian. 
\ack
KLB is supported by a UK EPSRC CASE studentship from Hewlett Packard. VMK is funded by a UK Royal Society University Research Fellowship. WJM was supported in part by MEXT and FIRST in Japan. We would like to thank Tim Spiller for useful discussions on the BCS Hamiltonian and the qubus architecture. 

\section*{References}
\bibliographystyle{unsrt}
\bibliography{bibliography,bibextra}

\end{document}